\documentclass[aps,prl,twocolumn,groupedaddress]{revtex4-2}
\usepackage{graphicx,color}
\usepackage{amsmath}
\usepackage{amssymb}
\usepackage{bm}
\usepackage{braket}
\usepackage{here}
\usepackage{comment}

\begin{document}


\title{Quantum droplet of a two-component Bose gas in an optical lattice near the Mott insulator transition}


\author{Yoshihiro Machida$^1$, Ippei Danshita$^1$, Daisuke Yamamoto$^2$, Kenichi Kasamatsu$^1$}
\affiliation{$^1$Department of Physics, Kindai University, Higashi-Osaka, Osaka 577-8502, Japan \\
$^2$Department of Physics, Nihon University, Sakurajosui, Setagaya, Tokyo 156-8550, Japan}


\date{\today}

\begin{abstract}
We theoretically study dynamical formation of a quantum droplet in a two-component Bose-Hubbard system with an external trap potential. Specifically, the superfluid in the central region surrounded by the Mott insulator with double filling forms a quantum droplet, which is self-bound thanks to the discontinuous nature of the quantum phase transition between the two phases. 
We show how to induce the characteristic behavior of the droplet through the control of the trap potential by using the time-dependent Gutzwiller simulations in a two-dimensional system. 
The static and dynamical properties of the droplet can be described qualitatively by the effective Ginzburg-Landau field theory with cubic-quintic nonlinearities, where the attractive cubic nonlinearlity emerges although all the bare interparticle interactions are repulsive. 
\end{abstract}


\maketitle


In the past few years, there has been tremendous interest in studying the properties of quantum droplets in ultracold gases \cite{ferrier2019ultradilute,luo2021new}. 
The droplet state is sustained by the energy competition between the well-tuned mean-field interactions and the intrinsic quantum fluctuations \cite{Petrov:2015}. 
Experimentally, a lot of efforts have been underway to reveal the properties of the quantum droplets \cite{Ferrier:2016Observation,Chomaz:2016Quantum,Semeghini:2018Self,cabrera2018quantum,Chomaz:2016Quantum,ferioli2019collisions}. 
While the previous experimental studies on quantum droplets have addressed only systems in continuum, recent theoretical studies have found droplet phases in a lattice system, which is specifically two-component Bose gases in an optical lattice (OL) with the repulsive intracomponent and attractive intercomponent interactions \cite{morera2020quantum,morera2021universal}. 
Ultracold atomic gases in an OL are nowadays standard platforms to implement versatile quantum simulators of quantum many-body systems \cite{bloch2008many,lewenstein2012ultracold}. 
The controllability of the system parameters from a weakly-interacting regime to a strongly-interacting one enables us to explore various states of matter. 

In this work, we propose an alternative mechanism to realize a quantum droplet in a two-component Bose-Hubbard (BH) system with both intra- and inter-component interactions are \textit{repulsive}. 
The Bose-Bose mixture in an OL is experimentally accessible \cite{catani2008degenerate,gadway2010superfluidity}. 
The previous theoretical studies have considered the ground state phase diagrams of the two-component BH model, predicting the existence of several phases and phase transitions; 
especially, there is the discontinuous (first-order) phase transition between the Mott insulator (MI) and the superfluid (SF) phase near the tip of the Mott-lobe in the spatial $d$-dimension with $d\geq 2$ \cite{kuklov2004commensurate,chen2010quantum,ozaki2012bose,yamamoto2013first,kato2014quantum,danshita2015cubic}. 
Also, nonequilibrium dynamics subject to the two-component BH model has been studied to reveal the impact of the inter-atomic interactions on dynamical behavior \cite{wernsdorfer2010lattice,hu2011detecting,zhang2012strong,lundh2012kelvin,shrestha2012fragmentation}.

Our dynamical simulations based on the Gutzwiller approximation show that the two-component Bose gas in a combined OL and trap potential can accommodate a localized structure of the SF component, provided by the discontinuous nature of the SF-MI transition. 
It is well known that when the interaction dominates over the kinetic energy, the density distribution of BH systems with a parabolic trap forms an inhomogeneous wedding-cake structure \cite{jaksch1998cold}. 
We find that when the transition between the central SF and the surrounding MI is of the first-order, the central SF does not expand even after the trap around it is locally turned off, behaving as a self-bound droplet. 
We explain the mechanism of the self-binding from a viewpoint of the effective Ginzburg-Landau (GL) theory, where the SF order parameter around the transition point can be described qualitatively by the effective GL equation with the attractive-cubic and repulsive-quintic nonlinearity. 
This type of nonlinear equation can support the solution of a self-bound droplet \cite{pushkarov1979self,pego2002spectrally,xi2016droplet}. 
Since the effective attraction is induced by the fourth-order perturbation process of the hoppings from the atomic limit, the droplet formation originates purely from quantum nature. 
Using the effective theory, we study stationary density profiles and collective excitations of the droplet, which are key features in the experimental observation. 

We consider the BH Hamiltonian for the two-component Bose gases in a two-dimensional square OL and a trap potential, which is given by
\begin{align}
\hat{H}=&\sum_{\alpha=1,2}\biggl{[}-J_{\alpha} \sum_{\braket{i,j}}(\hat{a}^{\dagger}_{\alpha,j}\hat{a}_{\alpha,i}+\hat{a}^{\dagger}_{\alpha,i} \hat{a}_{\alpha,j})-\sum_{j} \mu_{\alpha,j} \hat{n}_{\alpha,j} \nonumber \\ 
&+\frac{U_{\alpha}}{2}\sum_{j}\hat{n}_{\alpha,j}(\hat{n}_{\alpha,j}-1)\biggl{]}+U_{12}\sum_{j}\hat{n}_{1,j}\hat{n}_{2,j}
\label{2BHM}
\end{align}
with the component index $\alpha=1,2$ and the site index $j=(j_x,j_y)$. Here, $\braket{i,j}$ represents the nearest neighbor sites, $J_{\alpha}$ the hopping coefficient, $\mu_{\alpha,j}$ the local chemical potential, $U_{\alpha}$ and $U_{12}$ the intracomponent on-site interaction and the intercomponent one, respectively. 
The operators $\hat{a}_{\alpha,j}$ $(\hat{a}_{\alpha,j}^{\dagger})$ and $\hat{n}_{\alpha,j} \equiv \hat{a}^{\dagger}_{\alpha,j}\hat{a}_{\alpha,j}$ are the annihilation (creation) operator and the number operator at site $j$, respectively. 

We assume below the symmetric choice of the parameters, namely $J_1 = J_2 \equiv J$, $\mu_{j,1} = \mu_{j,2} \equiv \mu_j$, and $U_1=U_2 \equiv U >0$. 
For example, when we choose the two hyperfine states of $^{87}{\rm Rb}$ atom, $|F=2,m_F = -1\rangle$ and $|1,1\rangle$, as the two components, in a standard optical lattice the equal hoppings hold and the equal intracomponent interactions approximately do ($U_1/U_2\simeq 0.95$ \cite{mertes2007nonequilibrium}. 
By assuming further that the trap potential is equal for the two components, the equal local chemical potentials approximately imply the equal numbers of particles. The parameter $U_{12}$ can be controlled with use of a magnetic Feshbach resonance \cite{tojo2010controlling}.

We calculate the ground state and time-dependent dynamics of the system obeying Eq.~\eqref{2BHM} through the Gutzwiller approximation for the many-body wave function \cite{jaksch1998cold,ozaki2012bose,wernsdorfer2010lattice,lundh2012kelvin}. 
The Gutzwiller variational wave function is given by 
\begin{equation}
\ket{\Psi_\text{G}(t)}=\prod_{j}\sum_{n_{1},n_{2}=0}f_{n_{1},n_{2}}^{(j)}(t)\ket{n_{1},n_{2}}_{j}
\end{equation}
where $\ket{n_{1},n_{2}}_{j}$ represents the Fock state associated with the particle numbers for the two components at site $j$. 
The probability amplitude $f_{n_{1},n_{2}}^{(j)}$ satisfies the normalization condition $\sum_{n_{1},n_{2}}|f_{n_{1},n_{2}}^{(j)}|^{2}=1$. 
Under the variational principle, the minimization of $\bra{\Psi_\text{G}}\hat{H}-i\hbar \frac{d}{dt}\ket{\Psi_\text{G}}$ with respect to $f_{n_1,n_2}^{(j)*}$ gives the time-dependent equations for $f_{n_{1},n_{2}}^{(j)}$ as 
\begin{align}
& i\hbar \frac{df_{n_{1},n_{2}}^{(j)}}{dt} = \sum_{\alpha=1,2}\biggl{[}\frac{U}{2}n_{\alpha,j}(n_{\alpha,j}-1)-\mu_j n_{\alpha,j}\biggl{]} f_{n_{1},n_{2}}^{(j)}   \nonumber \\ 
& -J\sum_{i \in N(j)}\biggl{(}\Psi_{1,i}\sqrt{n_{1,j}}f_{n_{1}-1,n_{2}}^{(j)}+\Psi_{1,i}^{*}\sqrt{n_{1,j}+1}f_{n_{1}+1,n_{2}}^{(j)}\biggl{)} \nonumber \\
& -J\sum_{i \in N(j)}\biggl{(}\Psi_{2,i}\sqrt{n_{2,j}}f_{n_{1},n_{2}-1}^{(j)}+\Psi_{2,i}^{*}\sqrt{n_{2,j}+1}f_{n_{1},n_{2}+1}^{(j)}\biggl{)} \nonumber \\
& + U_{12} n_{1,j}n_{2,j}  f_{n_{1},n_{2}}^{(j)} .
 \label{2GW}
\end{align}
Here, $N(j)$ represents the nearest neighbor sites of site $j$. 
The SF order parameter for each component is written by $\Psi_{1,j}=\sum_{n_{1},n_{2}}f_{n_{1}-1,n_{2}}^{(j)*}\sqrt{n_{1}}f_{n_{1},n_{2}}^{(j)}$ and $\Psi_{2,j}=\sum_{n_{1},n_{2}}f_{n_{1},n_{2}-1}^{(j)*}\sqrt{n_{2}}f_{n_{1},n_{2}}^{(j)}$. 
In the following, real-time dynamics are simulated by solving Eq.~\eqref{2GW} with the Crank-Nicholson scheme, while the initial ground state is obtained by the imaginary time propagation of Eq.~\eqref{2GW}. 

\begin{figure}[H]
\centering
\includegraphics[width=1.0\linewidth]{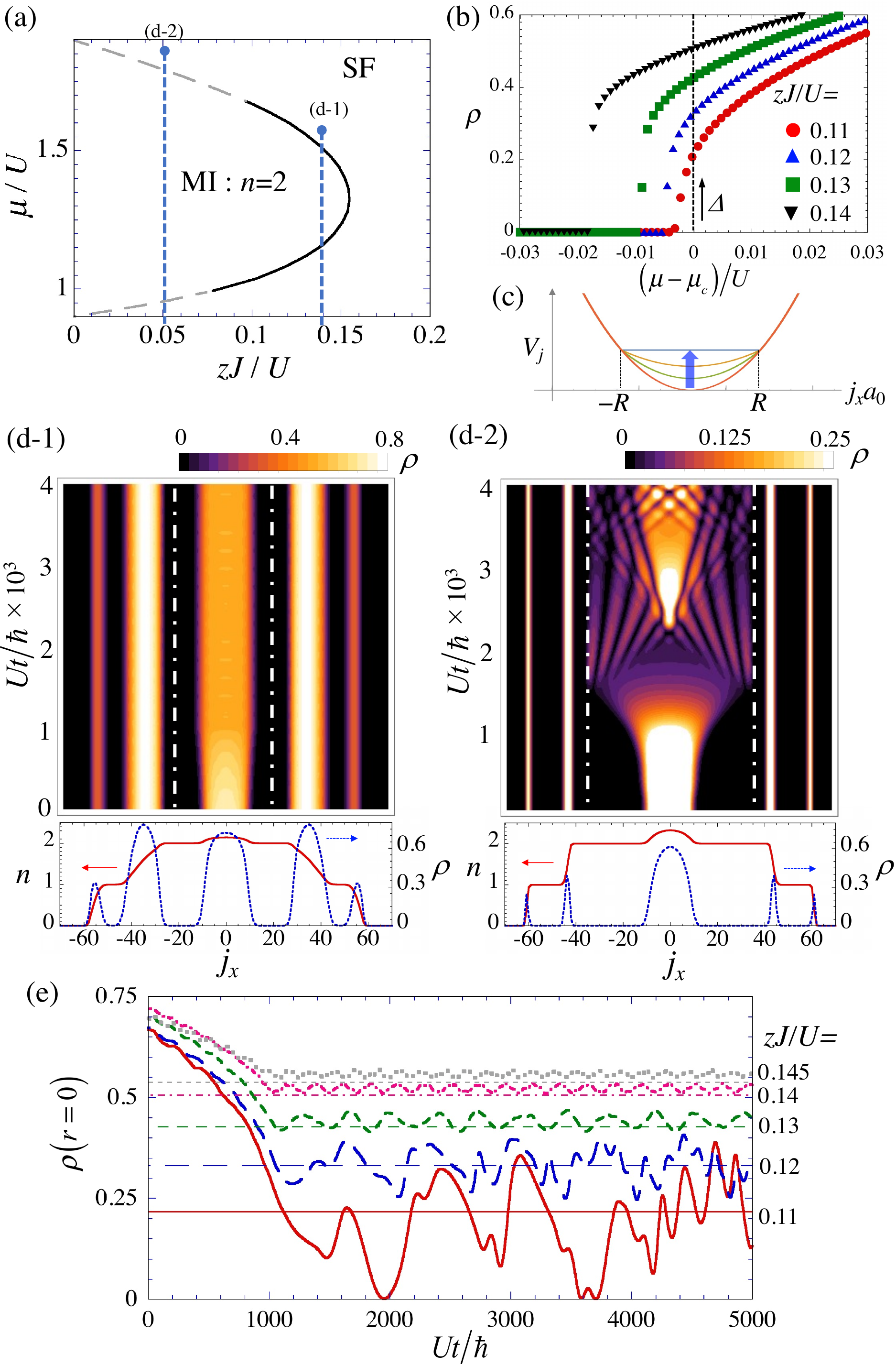} 
\caption{The panel (a) shows the ground state phase diagram in the $zJ/U$-$\mu/U$ plane obtained by the Gutzwiller analysis of the BH model Eq.~\eqref{2BHM} \cite{ozaki2012bose,yamamoto2013first}. 
The solid and dashed curves between SF and MI represent the first and second order transitions, respectively. 
The vertical dashed lines show the radial dependence of the equilibrium phase in the bottom panels of (d), determined by the local chemical potential. 
The panel (b) shows the equilibrium condensate density $\rho=\sum_\alpha |\Psi_{\alpha}|^{2}$ for several values of $zJ/U$ as a function of the chemical potential $\mu$ measured from the transition point $\mu_c$. 
For $\mu<\mu_c$ the SF state is metastable. 
In (c), we show how to change the trap potential for inducing characteristic behavior of a droplet formed by the SF in the central region of the trap.
The panels (d) show the time evolution of the cross section of the SF density along the $x$-axis starting from the initial states of the bottom panels, in which the profiles of the total particle density $n_1+n_2$ (red solid curve) and the SF density $\rho$ (blue dashed curve) are shown. 
The parameters are $U_{12}/U = 0.9$, $\tilde{k}=0.001$, and $(zJ/U, \mu/U, R)=(0.14, 1.56, 21)$ for (d-1) and $(0.05, 1.85, 35)$ for (d-2). 
Here, $\mu$ is chosen as $\mu=\mu_c+0.05U$
The vertical dashed-dotted lines show the radius $j_x = \pm R$ of the flat potential area. 
The panel (e) shows the time evolution of the condensate density at the center for several values of $zJ/U$. 
The horizontal thin lines show the gap $\Delta$ of the condensate density at the transition points for the corresponding $zJ/U$, obtained in (b). }
\label{mainfig}
\end{figure}
The mean-field phase diagram of the ground state in the homogeneous system is obtained in Refs.~\cite{ozaki2012bose,yamamoto2013first}, showing rich phase structures depending on the values of $U_{12}/U$. 
In this work, we confine ourselves to $U_{12}/U=0.9$;
the phase diagram in the $(zJ/U)$-$(\mu/U)$ plane is shown in Fig.~\ref{mainfig}(a), where $z=4$ is the coordinate number. 
The mean-field result is in qualitative agreement with the quantum Monte-Carlo result \cite{kato2014quantum}. 
There appear the two tricritical points (TCPs) for $0.68<U_{12}/U<1$ upon the boundary between the SF and the MI with double filling factors. 
In Fig.~\ref{mainfig}(b), the condensate density $\rho=\sum_\alpha |\Psi_\alpha|^2$ obtained from Eq.~\eqref{2GW} is plotted for several values of $zJ/U$ as a function of the chemical potential measured from the transition point, which is determined at the crossing point of the energy $\langle \Psi_\text{G} | \hat{H} | \Psi_\text{G} \rangle$ of the MI and that of the SF. 
At the transition point, the condensate density at the ground state shows a jump, denoted as $\Delta$, a clear signature of the first-order transition.

Next, we introduce the trap potential $V_j$. 
In the local density approximation, the chemical potential consists of the global value $\mu$ and the parabolic trap $V_j$ as $\mu_j = \mu - V_j = \mu - k (a_0 j)^2/2 $ with the spring constant $k$, the lattice constant $a_0$, and $j^2 = j_x^2 +j_y^2$.
Then, the equilibrium number distribution constitutes a wedding-cake structure depending on the global value of $\mu$. 
We choose $\mu/U$ such that the SF is positioned at the central region of the trap and the MI with $n=2$ surrounds the SF. 
The Thomas-Fermi radius of the density profile is given by $a_0 |j| =\sqrt{2\mu/k}$. 
The bottom panel of Fig.~\ref{mainfig}(d-1) shows the equilibrium profiles of the particle number $n$ and the SF density $\rho$ when 
the phase boundary between the central SF and the surrounding MI with $n=2$ associates with the first-order transition. 
In contrast, that of Fig.~\ref{mainfig}(d-2) shows the case when the phase boundary corresponds to the second-order transition. 

In the former case, the central SF can possess a characteristics of a droplet in the sense that the localized structure is dynamically kept even after the trap potential is turned off. 
To demonstrate this, we slowly change the potential $V_j$ under the protocol shown in Fig.~\ref{mainfig}(c) so as to flatten the area $j^{2} a_0^2 \leq  R^{2}$. 
Here, the radius $R$ of the flat area is determined at the position sufficiently inside the $n=2$ MI domain. 
The time sequence of the trap potential for $j^2a_0^2\leq R^2$ 
is set as 
$V_j = \delta\epsilon(t)(V_j^{(0)} - \epsilon_0) + \epsilon_0$, where 
$V_j^{(0)} = k(a_0j)^2/2$ and we change the time-dependent parameter as $\delta \epsilon(t) = 1 \to 0$ linearly within the interval $t=[0,T=1000\hbar/U]$. In the flat area, for $t>T$ the chemical potential takes a constant $\mu - \epsilon_0$, which is set to be the value inside the $n=2$ MI region.
This local manipulation of the trap is necessary since the SF droplet is sustained only in the presence of the surrounding MI domain. 
Such local control of the external potential can be achieved with recent experimental techniques, e.g., using the degital micromirror device \cite{mazurenko2017cold}.

We calculate the real time dynamics of the SF component using the time-dependent Gutzwiller approximation. 
The top panel of Fig.~\ref{mainfig}(d-1) represents the time evolution of $\rho$ starting from the initial state of the bottom panel ($zJ/U=0.14$), where we choose $R=21a_0$.  
Even though the confining potential is flat, the central SF component does not expand. 
We also show the time evolution of $\rho$ at the central site in Fig.~\ref{mainfig}(e).
After $U t / \hbar = 1000$, the central density makes an oscillation around a certain steady value. 
This steady value is almost coincident with the jump $\Delta$ of the condensate density at the first-order transition point. 
We confirmed that this localized structure appears when the central SF is surrounded by the $n=2$ MI through the first-order transition line.
When the simulation starts from the bottom panel of Fig.~\ref{mainfig}(d-2)  ($zJ/U=0.05$), on the other hands, the SF density expands and reaches to the edge of the flat region; the density makes an interference pattern due to the reflection from the potential edge. 

\begin{figure}[ht]
\centering
\includegraphics[width=1.0\linewidth]{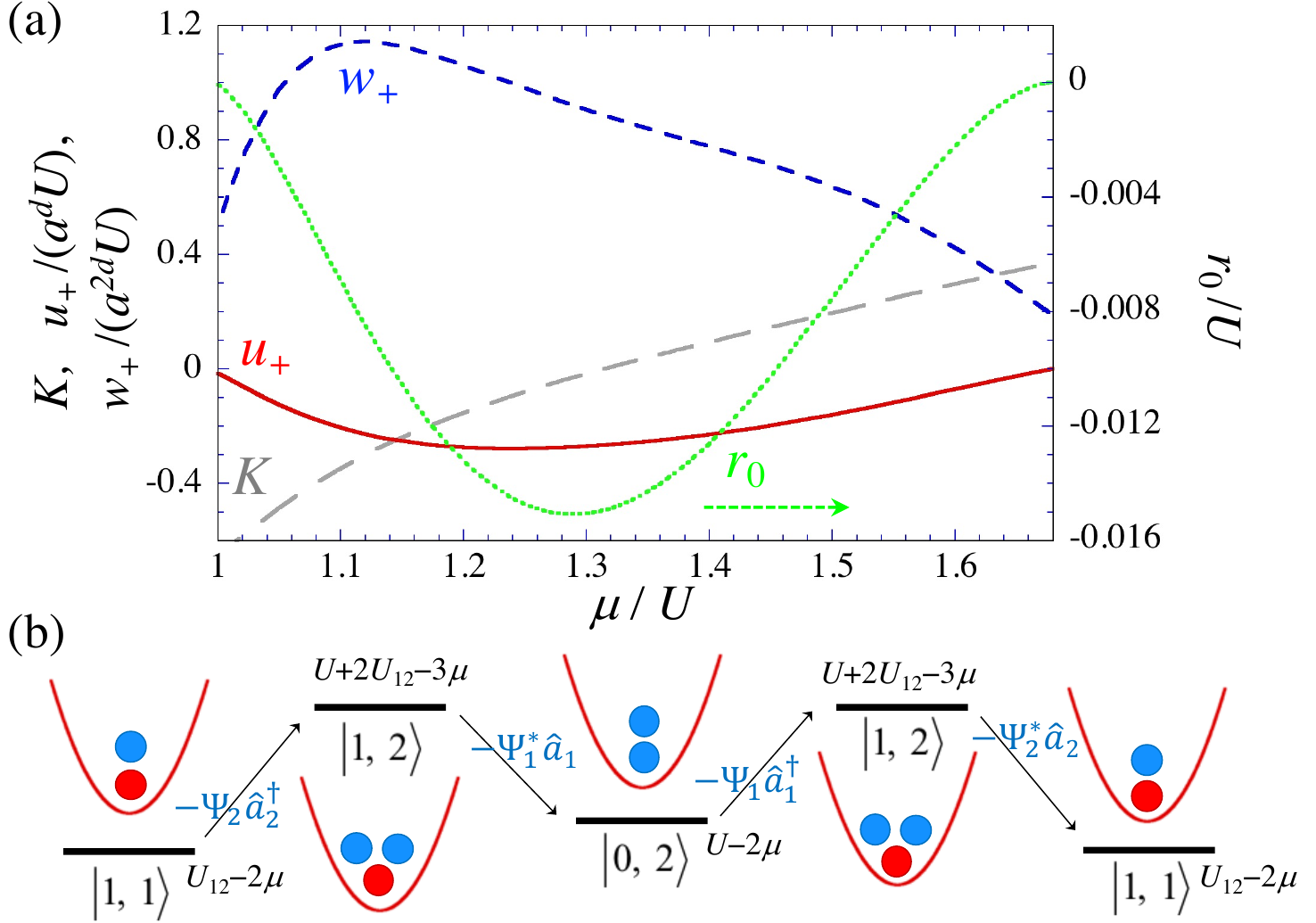} 
\caption{(a) The GL parameters along the first-order transition line between the SF and the $n=2$ MI as a function of $\mu/U$ for $U_{12}/U =0.9$. 
The solid, short dashed, and long dashed curves represent $u_+/(a_0^{d}U)$, $w_+/(a_0^{2d}U)$, and $K$, respectively (Here, $d=2$). 
The densely dotted curve represents $r_0/U$, whose vertical axis is the right side. 
(b) One of the fourth-order processes that contribute to the negativity of $u_{+}$. }
\label{glparameter}
\end{figure}

The appearance of the localized structure can be interpreted as the nature of the phase transition. 
For the first-order transition, the condensate density cannot continuously change down to zero at the SF-MI boundary. 
Thus, the monotonic expansion of the central density is prohibited so that the mean density cannot decrease below the gap $\Delta$. 
To understand these properties more clearly, we employ the effective sixth-order GL action expanded by the SF order parameter $\Psi_\alpha$ \cite{kato2014quantum,danshita2015cubic}. 
Assuming the symmetric form $\Psi_1 = \Psi_2 = \Psi$ for simplicity, we get the cubic-quintic GL equation
\begin{equation}
i\hbar K \frac{\partial \Psi}{\partial t}=\biggl{[}-\frac{\hbar^{2}}{2m}\nabla^{2}-r_0+u_{+}|\Psi|^{2}+w_{+}|\Psi|^{4}\biggl{]}\Psi \label{71}.
\end{equation}
Since we confine ourselves to the low-energy dynamics of the quantum droplet, we take only the first-order time derivative in the GL action. 
The effective mass is given by $m=\hbar^2/(2Ja_0^2)$.
Although the GL parameters have the spatial dependence since $\mu_j$ depends on the position in a trap, we approximate them by the values upon the transition line between the $n=2$ MI and the SF. 
The GL parameters $(K,r_0,u_+,w_+)$ along the first-order transition line is shown in Fig.~\ref{glparameter}(a). 
The negative $r_0$ ensures that the free energy around the MI phase takes a minimum. 

According to the GL theory, the emergence of the first-order transition requires $u_{+} < 0$, which is satisfied in a regime where the central SF forms the droplet, as seen in Fig.~\ref{glparameter}(a).
The negativity of $u_{+}$ can be understood by the fourth-order perturbation processes \cite{supple}, in which the hopping terms are regarded as the perturbation. 
One of the perturbation processes is shown in Fig.~\ref{glparameter}(b). 
When $U \sim U_{12}$, the energy of $|1,1 \rangle$ is nearly equal to that of the intermediate state $|2,0 \rangle$ or $|0,2 \rangle$.
Then, these processes give rise to an enhanced negative contribution of the coupling constant in front of $|\Psi_1|^2 |\Psi_2|^2$, which results in the attractive cubic-nonlinearlity in Eq.~\eqref{71}. 
Indeed, the first-order transition emerges only when $U \sim U_{12}$ (more precisely when $0.68<U/U_{12}<1$) according to the Gutzwiller analysis \cite{ozaki2012bose,yamamoto2013first}. 

The negative $u_{+}$ also provides a clear evidence of the existence of a self-bound droplet. 
First, let us see the stationary solution of Eq.~\eqref{71} by assuming the isotropy of the solution $\Psi(\bm{r}) = \psi(r)$. 
The solutions can be obtained through the imaginary time propagation of Eq.~\eqref{71} with the fixed norm $\int d^2r |\psi|^2 \equiv N_\rho$, where $r_0$ reads the chemical potential of the condensate. 
For $u_+ < 0$ we can get localized solutions even in a free space \cite{pushkarov1979self,pego2002spectrally,xi2016droplet}. 
With increasing $N_\rho$, the solution exhibits a ``flat-top" shape, which is typical of the droplet structure \cite{luo2021new,Petrov:2015}. 
When the droplet size is much larger than the length scale $l_{w} \equiv \hbar/\sqrt{m w_+ n_0^2}$, where $n_0$ is the density at the flat region of the droplet, the derivative term $(1/r) \partial_r$ in the Laplacian can be neglected. 
Then, the droplet solution with the radius $R_\text{d}$, at which the density becomes a half of that of the center, can be written by \cite{danshita2015cubic}
\begin{equation}
\psi^2 = \frac{n_0}{2} \left[ 1- \tanh\left( \sqrt{\frac{2}{3}} \frac{r-R_\text{d}}{l_w} \right) \right].   \label{danshitaexactsol}
\end{equation}
Here, $n_0$ is given by the jump $\Delta_\text{GL}= - 3 u_+ /(4w_+)$ of the SF density at the first-order transition point. 
The numerical solution of Eq.~\eqref{71} can be well approximated by Eq.~\eqref{danshitaexactsol}, the central density being slightly exceeded from the above estimation $2n_0$ \cite{supple}. 

\begin{figure}[ht]
\centering
\includegraphics[width=1.0\linewidth]{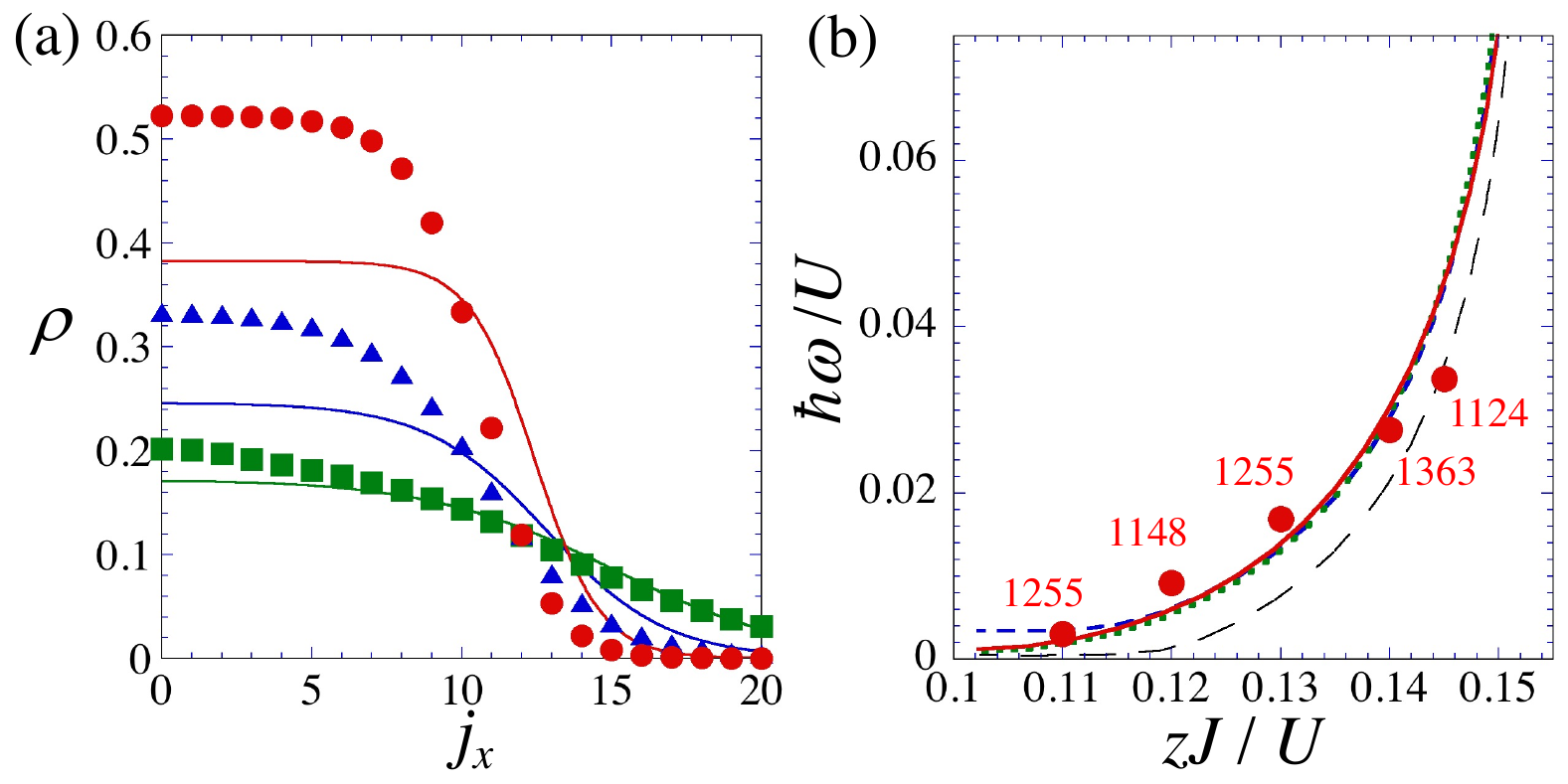} 
\caption{Comparison between the Gutzwiller calculation and the GL analysis. 
In (a), the red, blue, green curves represent the stationary profiles of the condensate density $\rho = 2 |\psi|^2$ by solving Eq.~\eqref{71} for the GL parameters corresponding to $(zJ/U,\mu/U) =$ (0.14,1.51), (0.12,1.61), and (0.11.1.64), respectively. 
The profiles obtained from the dynamical Gutzwiller simulation are also shown by circles, triangles, and squares for the same parameters. 
The panel (b) shows the eigenfrequencies of the collective excitations with the quantum numbers $(n,l)=(1,0)$, obtained by solving the BdG equation for $N_\rho (2U/zJ) \equiv N' = 100$ (long-dashed), 200 (short-dashed), 500 (dotted) and 1000 (solid). 
The circles represent the results by the Gutzwiller simulations, where the frequency is extracted from the power spectrum of the oscillation in Fig.~\ref{mainfig}(e) and the attached number represents $N'$. }
\label{bdgspect}
\end{figure}
In Fig.~\ref{bdgspect}(a), we plot the typical stationary solutions of Eq. \eqref{71} as well as the profile of the condensate density obtained from the Gutwiller simulations in Fig.~\ref{mainfig}, averaged for $3000\leq Ut/\hbar \leq 5000$. 
Here, the norm of $\psi$ is determined to reproduce the condensate number $N_\rho$ within the droplet of $| \Psi_\text{G} \rangle$ \cite{supple}.
The profiles of the Gutzwiller calculations are well described by the GL results as the parameters approach to the TCP. 
This is naturally understood that the GL expansion is quantitatively validated for a smaller values of a jump of the SF order parameter at the transition point. 
Nevertheless, the overall qualitative feature of the quantum droplet is well captured by the GL solutions. 
For example, the central density of the Gutzwiller calculation is determined by the density jump $\Delta$ shown in Fig.~\ref{mainfig}(b), whose property has been clearly seen in Fig.~\ref{mainfig}(e).  

The small oscillation seen in Fig.~\ref{mainfig}(e) is related with the collective modes of the quantum droplet \cite{tylutki2020collective,hu2020collective,sturmer2021breathing} induced by the temporal change of the potential. 
We calculate the frequencies of the collective excitations by the Bogoliubov-de Genne (BdG) analysis, in which we expand the order parameter as $\Psi(\bm{r},t)=[\Psi(\bm{r})+\delta \Psi(\bm{r},t)]e^{-i r_0 t/\hbar}$ with $\delta \Psi(\bm{r},t)=\sum_{n} [ u_{n}(r)e^{il \theta -i \omega t}-v_{n}^{*}(r)e^{-il \theta +i\omega t} ] $ and solve the eigenvalue equations.
Figure \ref{bdgspect} (b) shows the excitation frequencies with $(n,l)=(1,0)$, i.e., breathing mode, along the first-order transition line as a function of $zJ/U$. 
The frequency is decreased from the tip of the Mott lobe ($zJ/U \simeq 0.155$) to the TCP ($zJ/U \simeq 0.096$), being insensitive to the norm for $N_\rho (2U/zJ) > 200$. 
We also plot the oscillation frequencies, obtained from the main peak of the power spectrum, of the central density in the Gutzwiller simulations [Fig.~\ref{mainfig}(e)]. 
The plots are reasonably coincident with the frequencies of the BdG analysis, since excitations with $l \neq 0$ cannot be induced in our potential protocol. 
We note that since a number of modes with $n \geq 2$ are lying above the $n =1$ mode with small energy gaps especially near the TCP, multiple modes are excited simultaneously.
Although the GL approximation should become better near the TCP, it is difficult to make a quantitative comparison of the BdG analysis since a pure breathing oscillation with a small amplitude cannot occur there. 

In summary, we have shown that a suitable manipulation of an external trap leads to the formation of a quantum droplet of the SF phase surrounded by the $n=2$ MI in a two-component BH system.  
The underlying mechanism of the droplet formation is the effective attraction between the intercomponent SF order parameters, which originates from the fact that the fourth-order process of the perturbative expansion is enhanced for $U_{12} \sim U$, even though the on-site interactions are all repulsive. 
We showed through the time-dependent Gutzwiller analysis that the localized structure of the SF phase can be kept even after the external trap is locally flattened. 
The static and dynamical properties of the droplet can be described qualitatively by the effective GL theory with the cubic-quintic nonlinearity. 

This work was financially supported by JSPS KAKENHI
Grants Nos.\ JP18K03472 (K.K.), JP18H05228 (I.D.), JP21H01014 (I.D.), JP18K03525 (D.Y.) and JP21H05185 (D.Y.),
by JST CREST Grant No.\ JPMJCR1673 (I.D. and D.Y.), by JST FOREST Grant No.\ JPMJFR202T (I.D.),
and by MEXT Q-LEAP Grant No.\ JPMXS0118069021 (I.D.).

\bibliography{reference}

\newpage
\begin{widetext}

\begin{center}
\textbf{Supplemental Material for ``Quantum droplet of a two-component Bose gas in an optical lattice near the Mott insulator transition"}
\end{center}

\subsection{The GL action and the coupling constant $u_{12}$}
The derivation of the effective sixth-order GL action is described in Ref.~\cite{kato2014quantum}, the form of the action being 
\begin{align}
S^{\rm{eff}} =\int dt \int d^{d}x \Biggl[ \sum_{\alpha=1,2}\biggl( & i \hbar K \Psi_{\alpha}^{*}\frac{\partial \Psi_{\alpha}}{\partial t}-\hbar^{2} W \biggl{|}\frac{\partial \Psi_{\alpha}}{\partial t}\biggl{|}^{2}
 -\frac{\hbar^{2}}{2m}|\nabla \Psi_{\alpha}|^{2}+r_0 |\Psi_{\alpha}|^{2}-\frac{u}{2}|\Psi_{\alpha}|^{4}-\frac{w}{3}|\Psi_{\alpha}|^{6}\biggl) \nonumber \\
& -u_{12} |\Psi_{1}|^{2}|\Psi_{2}|^{2}-w_{12} \left( |\Psi_{1}|^{4}|\Psi_{2}|^{2}+|\Psi_{1}|^{2}|\Psi_{2}|^{4} \right) \Biggl]. \label{6gl}
\end{align}
The GL parameters are given as a function of the parameters $(zJ/U, \: U_{12}/U, \: \mu_\alpha/U)$ of the BH Hamiltonian 
\begin{align}
\hat{H}=\sum_{\alpha=1,2}\biggl{[}-J \sum_{\braket{i,j}}(\hat{a}^{\dagger}_{\alpha,j}\hat{a}_{\alpha,i}+\hat{a}^{\dagger}_{\alpha,i} \hat{a}_{\alpha,j}) - \mu_{\alpha} \sum_{j} \hat{n}_{\alpha,j}  +\frac{U}{2}\sum_{j}\hat{n}_{\alpha,j}(\hat{n}_{\alpha,j}-1)\biggl{]}+U_{12}\sum_{j}\hat{n}_{1,j}\hat{n}_{2,j}
\label{2BHM}
\end{align}
through a perturbative expansion, the explicit form being given in Ref.~\cite{kato2014quantum}. 
Here, we give only the expression of the expansion parameter $u_{12}$, which is important in our study since its negativity realizes the first-order transition. 

Let us treat the hopping term in the BH Hamiltonian as a perturbation. 
The ground state of the unperturbed Hamiltonian is given by a product of local Fock state $| n_1,n_2 \rangle = | g,g \rangle$ and the unperturbed energy is 
\begin{equation}
E_{n_1,n_2} = \sum_\alpha \left[ -\mu_\alpha n_\alpha + \frac{U}{2} n_\alpha (n_\alpha -1 )\right] + U_{12} n_1 n_2. 
\end{equation}
In our work, we have focused on $g=1$, namely $n=n_1+n_2=2$.
Using this, the fourth-order perturbation analysis yields the expansion parameter $u_{12}$ of the GL action as
\begin{align}
u_{12} = a_0^d Z^4 J^4 & \Biggl[  \left( \frac{g+1}{E_{g+1,g} - E_{g,g}} + \frac{g}{E_{g-1,g} - E_{g,g}}    \right) \left( \frac{g+1}{(E_{g,g+1} - E_{g,g})^2} + \frac{g}{(E_{g,g-1} - E_{g,g})^2}    \right)  \nonumber \\
&+\left( \frac{g+1}{E_{g,g+1} - E_{g,g}} + \frac{g}{E_{g,g-1} - E_{g,g}}    \right) \left( \frac{g+1}{(E_{g+1,g} - E_{g,g})^2} + \frac{g}{(E_{g-1,g} - E_{g,g})^2}    \right)  \nonumber \\ 
&-\left( \frac{1}{E_{g+1,g} - E_{g,g}} + \frac{1}{E_{g,g+1} - E_{g,g}}  \right)^2 \frac{(g+1)^2}{E_{g+1,g+1} - E_{g,g}} \nonumber \\
&-\left( \frac{1}{E_{g+1,g} - E_{g,g}} + \frac{1}{E_{g,g-1} - E_{g,g}}  \right)^2 \frac{g(g+1)}{E_{g+1,g-1} - E_{g,g}} \nonumber \\
&-\left( \frac{1}{E_{g-1,g} - E_{g,g}} + \frac{1}{E_{g,g+1} - E_{g,g}}  \right)^2 \frac{g(g+1)}{E_{g-1,g+1} - E_{g,g}} \nonumber \\
&-\left( \frac{1}{E_{g-1,g} - E_{g,g}} + \frac{1}{E_{g,g-1} - E_{g,g}}  \right)^2 \frac{g^2}{E_{g-1,g-1} - E_{g,g}} \Biggr]. 
\label{u12peert}
\end{align}
When $U \sim U_{12}$, the energy differences $E_{g+1,g-1} - E_{g,g}$ and $E_{g-1,g+1} - E_{g,g}$ in the fourth and fifth lines in Eq.~\eqref{u12peert} become small, so that these negative contributions are resonantly enhanced.

\subsection{The GL analysis in a uniform system}
The GL equation 
\begin{equation}
i\hbar K \frac{\partial \Psi}{\partial t}=\biggl{[}-\frac{\hbar^{2}}{2m}\nabla^{2}-r_0+u_{+}|\Psi|^{2}+w_{+}|\Psi|^{4}\biggl{]}\Psi \label{71appl}.
\end{equation}
is useful to understand the SF-MI transition of the two-component BH model as discussed thoroughly in Refs.\cite{kato2014quantum,danshita2015cubic}.
Here, we summarize several simple results obtained from the stationary GL equation in a uniform system to understand the present work. 

First, let us consider a uniform solution $\Psi(\bm{r}) = \sqrt{n_0}$ of the GL equation, where assume $\Psi_1=\Psi_2=\Psi$. 
We also suppose $w_+ > 0$ below to ensure the stability of the solution. 
Equation~\eqref{71appl} then reads 
\begin{equation}
\left( -r_0 + u_+ n_0 + w_+ n_0^2 \right) \sqrt{n_0} = 0,   \label{statGL}
\end{equation}
and yields three solutions
\begin{align}
n_0 = 0, \quad \frac{-u_+ + \sqrt{D}}{2 w_+}, \quad \frac{-u_+ - \sqrt{D}}{2 w_+}  \label{solgl}
\end{align} 
with $D=u_+^2 + 4 r_0 w_+$. 

When $u_+ \geq 0$, the first two solutions of Eq.~\eqref{solgl} can satisfy $n_0 \geq 0$. For $r_0 < 0$, the physical solution is only $n_0=0$, which means that the ground state is the MI. 
For $r_0 > 0$, $n_0$ can have a positive finite value and the SF becomes the ground state.  
Thus, the transition takes place at $r_0 = 0$ and is of the second order type. 

\begin{figure}[ht]
\centering
\includegraphics[width=0.4\linewidth]{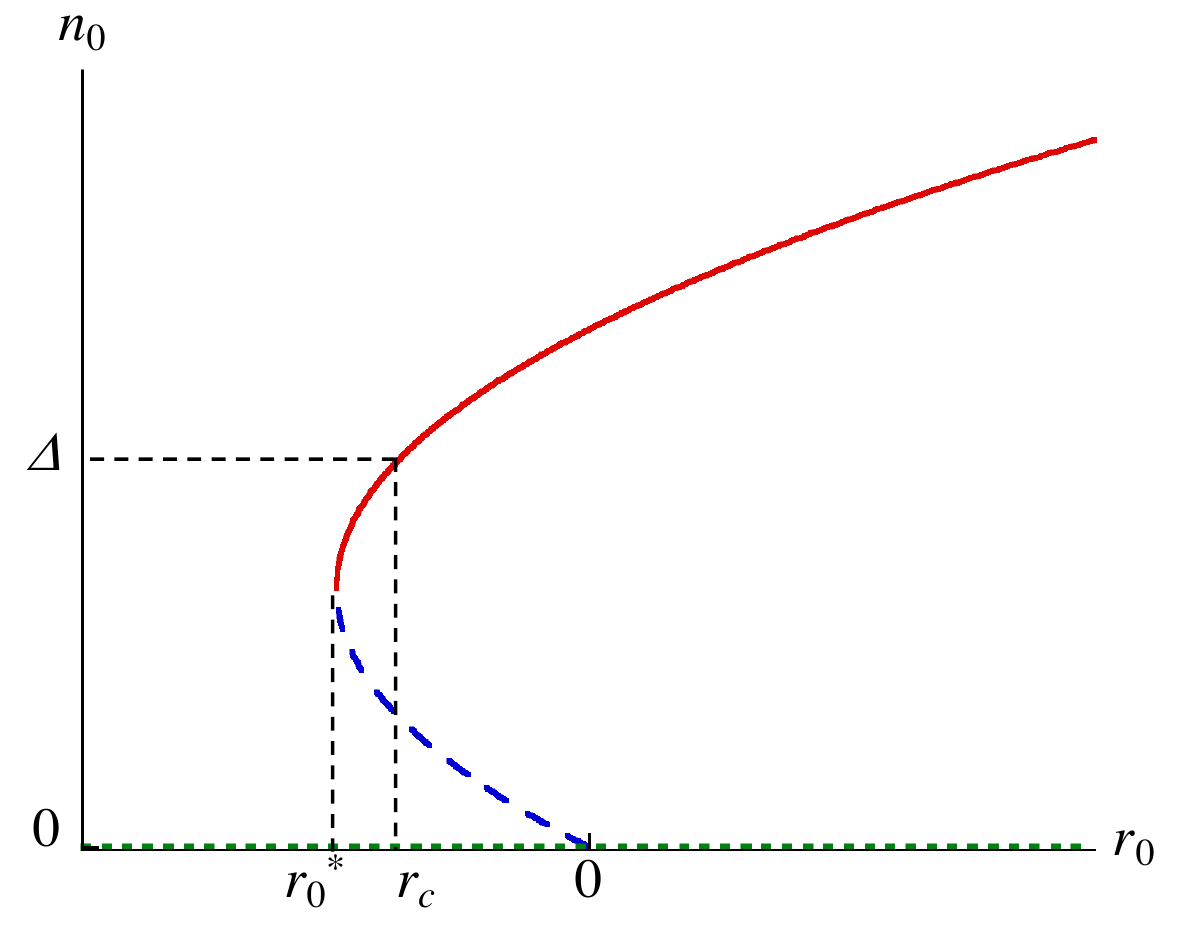} 
\caption{The three solutions of Eq.~\eqref{solgl} as a function of $r_0$, where the dotted line, the solid red curve, and the dashed blue curve represent the 1st, 2nd and 3rd solutions, respectively. }
\label{app1}
\end{figure}
When $u_+ < 0$, all solutions of Eq.~\eqref{solgl} can satisfy $n_0 \geq 0$. 
The three solutions are depicted in Fig.~\eqref{app1} as a function of $r_0$.
The second solution becomes a positive finite value when $D>0$, namely $r_0 \geq -u_+^2/(4w_+) \equiv r_0^{\ast}$, being stable branch satisfying $\partial n_0/ \partial r_0 > 0$. 
The last solution becomes positive finite when $r_0^{\ast} \leq r_0 \leq 0$, corresponding to an unstable branch with the relation $\partial n_0/ \partial r_0 < 0$. 
The first-order transition point can be determined by means of Maxwell's construction as $r_0 = -3u_+^2/(16w_+) \equiv  r_c$. 
The MI state with $n_0=0$ is the ground state when $r_0 < r_c$, a metastable when $r_c < r_0 < 0$, and an energetically unstable when $r_0 >0$. 
The SF state with $n_0 \neq 0$ is the ground state when $r_0 > r_c$ and a metastable state when $r_{0}^\ast < r_0 < r_c$. 
Substituting $r_0 = r_c$ into the second solution of Eq.~\eqref{solgl}, we evaluate the jump $\Delta_\text{GL}$ of $n_0$ at the transition point as $\Delta_\text{GL} = -3u_+/(4w_+)$. 

\subsection{The dimensionless form of the GL equation}
In order to scale Eq.~\eqref{71appl} into the dimensionless form, we replace the order parameter as $\Psi \to \tilde{\Psi}/a_0$ with the lattice constant $a_0$. 
Dividing the both sides of Eq.~\eqref{71appl} by $U$ of the BH Hamiltonian, we have 
\begin{equation}
i\hbar \frac{K}{U} \frac{\partial \tilde{\Psi}}{\partial t}=\biggl{[}-\frac{\hbar^{2}}{2mU}\nabla^{2}-\frac{r_0}{U}+\frac{u_{+}}{U a_0^2}|\tilde{\Psi}|^{2}+\frac{w_{+}}{Ua_0^4}|\tilde{\Psi}|^{4}\biggl{]} \tilde{\Psi}. \label{71appl2}.
\end{equation}
Taking the characteristic time and length scale as $\tau = \hbar K /U$ and $\xi = \hbar /\sqrt{mU}$, we obtain the dimensionless form of the GL equation 
\begin{equation}
i \frac{\partial \tilde{\Psi}}{\partial \tilde{t}}=\biggl{[}-\frac{1}{2}\tilde{\nabla}^{2}-\tilde{r}_0+\tilde{u}_{+} |\tilde{\Psi}|^{2}+\tilde{w}_{+} |\tilde{\Psi}|^{4}\biggl{]} \tilde{\Psi}. \label{71appldimless}.
\end{equation}
Here, the normalization of the order parameter $N_\rho = \int d^2r |\Psi|^2$ in the dimensionless form is given by 
\begin{equation}
\int d^2 \tilde{r} |\tilde{\Psi}|^2 = N_\rho \frac{a_0^2}{\xi^2} \equiv N'.
\end{equation}
The factor $a_0/\xi$ is written as $a_0/\xi = \sqrt{2U/(zJ)}$. 
We numerically solve Eq.~\eqref{71appldimless} through the imaginary time propagation by fixing the norm $N'$ by assuming a radial isotropy $\Psi(\bm{r}) = \psi(r)$.
In Fig.~\ref{app2}, we show the solutions using the GL parameters $(u_+,w_+) = (-0.151,0.620)$ at $(zJ/U,\mu/U) = (0.14,1.51)$ for some typical values of $N'$.
\begin{figure}[ht]
\centering
\includegraphics[width=0.4\linewidth]{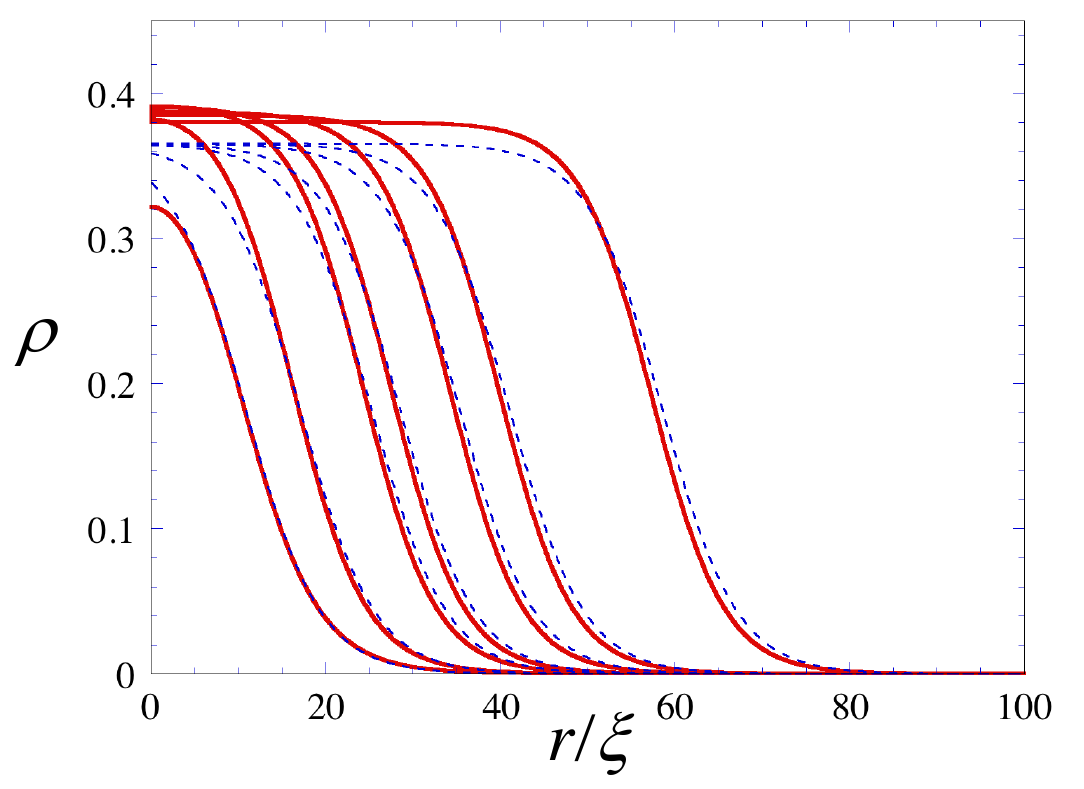} 
\caption{The Condensate density $\rho = 2 |\psi(r)|^2$ obtained from the solution of the GL equation for the GL parameters at $(zJ/U,\mu/U) = (0.14,1.51)$ and $N' = 100$, 200, 400, 500, 750, 1000, and 2000. The solid curves are numerical solutions, while the dashed curves are analytical solutions Eq.~(5) in the text. }
\label{app2}
\end{figure}

The Gaussian-like profile for a small $N'$ transforms to the flat-top shape with increasing $N'$. 
We also plot the approximate analytic solution given by Eq.(5) in the text, which can be obtained by neglecting the first-order derivative $\partial_r /r$ in the Laplacian. 
Here, the radius $R_\text{d}$ is determined from the norm $N'$. 
The approximate solutions are in reasonably good agreement with the numerical ones, but the central density of the numerical solution is slightly exceeded from that of the analytic solution; the latter is exactly given by $2\Delta_\text{GL}$.

\end{widetext}

\end{document}